\documentclass[english,showpacs,floatfix,12pt]{revtex4}
\usepackage[dvips]{graphicx}
\usepackage{longtable}
\usepackage{amsmath,amssymb}
\usepackage{dcolumn}
\usepackage{latexsym}
\usepackage{babel}
\usepackage[latin1]{inputenc}
\usepackage{color}
\usepackage[colorlinks]{hyperref}
\newcommand\tabcaption{\def\@captype{table}\caption}

\def\dl{\delta}
\def\Dl{\Delta}

\def\wt{\widetilde}
\def\nn{\nonumber}

\def\l{\left}
\def\r{\right}

\begin{document}
\title{\LARGE \bf Test of Einstein's Mass-Energy Relation}
\author{Ying-Qiu Gu}
\email{yqgu@fudan.edu.cn} \affiliation{School of Mathematical
Science, Fudan University, Shanghai 200433, China} 
\date{2nd June 2017}

\begin{abstract}
The Einstein's mass-energy relation $E=mc^2$ is one of the most
fundamental formulae in physics, but it has not been seriously
tested by an elaborated experiment, and only some indirect evidences
in nuclear reaction suggested that it holds to high precision.
Manifestly, for a particle, different self potential leads to
different energy-speed relation, which can be used as the
fingerprints of them.  In this letter, we propose an experiment  to
test this relation. The experiment only involves low energy of
particles and measurement of speed, which can be easily realized.
The experiment may shed lights on a number of fundamental puzzles in
physics. \vskip0.3cm \noindent{\bf Keywords: } {\em mass-energy
relation, interactive potential, energy-speed relation}
\end{abstract}
\maketitle

In Einstein's original paper \cite{E1}, he derived the the kinetic
energy of a particle $K$,
\begin{eqnarray}K=mc^2-m_0 c^2,
\qquad m={m_0}\l(1-\frac {v^2}{c^2}\r)^{-\frac 1 2},
\label{1.1}\end{eqnarray} which implies the total energy and the
speed of a particle have the following relation
\begin{eqnarray}E=m(v) c^2. \label{1.2}\end{eqnarray}
However, the Einstein's derivation is based on the linear classical
mechanics, and this relation has not been directly tested by
elaborated experiment. There once some indirect evidences in the
nuclear reaction. The most accurate one is provided by S. Rainville
et al.\cite{test1}, which indicates that the mass-energy relation
$E=m c^2$ holds to an error level less than 0.00004\% in the process
of neutron capture by nuclei of sulfur and silicon resulting in
$\gamma$-radiation. As pointed out by E. Bakhoum\cite{test2}, it is
actually a test for the energy conversion $\Dl E=\Dl m c^2$ at low
speed of the particles.

Strange enough, as one of the most fundamental relation, a direct
test for the energy-speed relation (\ref{1.2}) is absent. It seems
to be interesting for few people, and one can hardly find a
literature involving the problem. As a matter of fact, this relation
is related with a lot of fundamental problems in physics, such as
the relationship between quantum mechanics and classical one, the
self-potentials of an elementary particles, the Lorentz
transformation law for different parameters and Lorentz violation,
the structure of the space-time etc. The main purpose of this letter
is to draw the attention of physical society towards this problem.

The Dirac equation with different potentials describes elementary
particles, and maybe give some explanations for dark matter and dark
energy\cite{spn1,spn2,spn3,spn4,spn5}.  For spinors with
self-potentials such as nonlinear potential and electromagnetic one
$A^\mu$, detailed calculation shows the potentials result in the
different energy-speed relation, which can be used as fingerprints
of the interactions. In classical mechanics, these relations are
concealed.  Taking $c=1$ as unit of speed, we find the general
representation of the energy-speed relation takes the following
form\cite{gu1,gu2},
\begin{eqnarray}
E(v)= \frac {M_0}{\sqrt{1-v^2}} -\frac{M_1
v^2}{\sqrt{1-v^2}}+\frac{W_F}{\sqrt{1-v^2}}
\ln\frac{1}{\sqrt{1-v^2}},\label{1.3}
\end{eqnarray}
where $(M_0,M_1,M_F)$ are all constants of mass dimension, and $M_0$
is the total static mass of the particles, which is the relativistic
effect reflecting the structure of the space-time. $M_1$ corresponds
to interactions such as electromagnetic potential. $M_F$ corresponds
to the nonlinear self-interactive potential. $M_1$ and $M_F$ reflect
the structure of a particle. For normal particles such as electron,
we have $M_F\sim 10|M_1|\ll  M_0$.

The nonlinear effects can be hardly detected under normal conditions
due to little values of $(M_1,M_F)$ and the function
$\ln\sqrt{1-v^2}$. For example, when an electron get kinetic energy
30MeV, the corresponding speed reaches $v_1=0.99986c$, but
\begin{eqnarray}
E(v_1)\dot=\frac{M_0-M_1+4 M_F}{\sqrt{1-v_1^2}}\approx
\frac{M_0}{\sqrt{1-v_1^2}}.\label{1.6*}
\end{eqnarray}
That is to say, (\ref{1.3}) is a stiff equation of the
coefficients $(M_0,M_1,M_F)$. So we have to develop some tricks to
get meaningful solution.

\begin{figure}
\centering
\includegraphics[width=16cm]{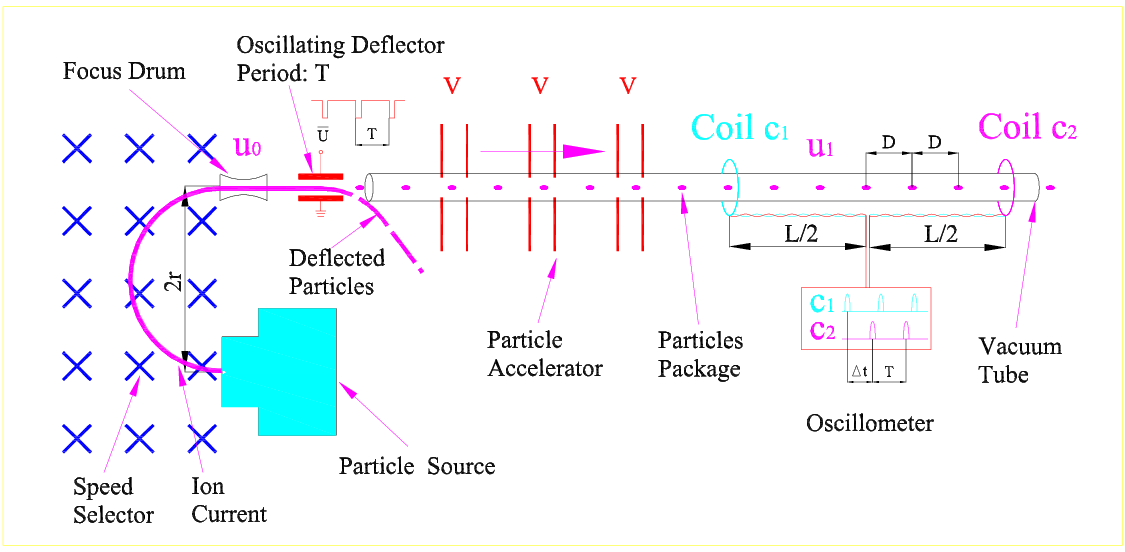}
\caption{The flow chart and experimental scheme to test the
mass-energy relation} \label{fig}
\end{figure}

In this letter, we propose the following experimental project, which
can sensitively measure the curve of the energy-speed relation. The
flow chart and experimental scheme are illustrated in Fig.\ref{fig}.

\begin{enumerate}
\item The particles with unit charge are produced by the particles source,
and the ones at initial speed $v_0$ are selected by a homogeneous
magnetic field. By adjusting the radius $r$, we can control the
initial speed  $v_0$ of the particles.

\item The particle beam with speed $v_0$ is focused by a focus drum.

\item The focused particle beam is controlled by an oscillating deflector with signal period $T$.
When the plate condenser is charged with voltage $\bar U$, the
particles are deflected and can not enter the vacuum tube. Only in
the moment the voltage $\bar U$ is removed, a package of charged
particle moves into the vacuum tube to be accelerated and tested.
Then we get a series of particle packages with periodic intervals
\begin{eqnarray}
v_0 T\to D=v_1 T. \label{t0}
\end{eqnarray}

\item The series-wound accelerator is constructed by a set of uniform
electrodes, which can be charged with high voltage $V$. When the
selected particles pass though one pair electrodes, each particle
receives an energy increment $\dl E=eV$, which converts into its
kinetic energy. If $n$ pair electrodes are charged, then we get the
total kinetic energy increment for each particle
\begin{eqnarray}
K=E(v_1)-E(v_0) = n\dl E =neV.\label{1.4}
\end{eqnarray}
(\ref{1.3}) and (\ref{1.4}) establish the connection between the
speed $v$ and $nV$.

\item An oscillometer with high frequency(e.g. 100MHz) is placed at the
midpoint between the two solenoid coils. When one particles package
pass trough one coil, an electromotive signal is generated in the
coil and the signal can by detected by the oscillometer. The time
increment between the package passing trough two solenoid coils
$c_0$ and $c_1$ is
\begin{eqnarray}
\Dl t=\frac L {v_1},  \label{t1}
\end{eqnarray}
which can be measured by the oscillometer.

\item Adjusting the deflecting period $T=\Dl t$ until we get two
series of periodic pulses which can be completely overlapped, then
by (\ref{t0}) and (\ref{t1}) the final speed $v_1$ of the particles
can be exactly calculated as
\begin{eqnarray}
v_1=\frac L T.  \label{t2}
\end{eqnarray}
If the distance between two coils $c_0$ and $c_1$ is set as $L=6m$,
then the deflecting frequency $\nu\sim 50$MHz or $T\sim 0.02\mu \rm
s$.

\item Since the distance $L$ and the period $T$ can be adjusted to very high accuracy,
so we can calculate the true energy-speed relation $E=m(v) c^2$
according to the measured results.

\end{enumerate}

In what follows, we take (\ref{1.3}) as example to show the dada
treatment, i.e. we determine the constants $(M_0,M_1,M_F)$ by
fitting the curve $f(v_1,V)=0$ defined by (\ref{1.3}). Now we make
some simplification of (\ref{1.4}). At first, we can solve the
static mass $M_0$ at low energy $v=\wt v_1\ll c$ as follows. Assume
the voltage $V=V_0$, in this nonrelativistic case, we have the
approximation of (\ref{1.4}) as follows
\begin{eqnarray}
neV_0 &\dot = &\frac 1 2 (M_0-2 M_1+M_F)(\wt
v_1^2-v_0^2).\label{1.5}
\end{eqnarray}
Then we get
\begin{eqnarray}\l\{\begin{array}{lll}
M_0&=&m_s+2 M_1-M_F, \\
E_0&=& m_s+2M_1-M_F+\frac 1 2 m_su_0^2\dot =E(v_0),\\
m_s&\equiv&2n e V_0(\wt v_1^2-v_0^2)^{-1},\end{array}\r. \label{1.6}
\end{eqnarray}
where $m_s$ is the nonrelativistic static mass of the particle in
classical sense. Therefore, we only need to determine two little
coefficients $(M_1,M_F)$ at high energy.

By (\ref{1.3}), (\ref{1.4}) and (\ref{1.6}), we have the following
relation
\begin{eqnarray}
(2-v_1^2)M_1&-&\l(1+\ln\sqrt{1-v_1^2}\r)M_F  \l\{=
E(v_1)\sqrt{1-v_1^2}-m_s \r.
\qquad {\rm (by ~(\ref{1.3})~ and~ (\ref{1.6}))}\nn\\
& = & \l. [neV+E(v_0)]\sqrt{1-v_1^2}-m_s \r\} \qquad\qquad
\qquad\qquad\quad~
{\rm (by~ (\ref{1.4})~ and~ (\ref{1.6}))}\nn\\
& \dot = & (neV+\frac 1 2 m_s v_0^2)\sqrt{1-v_1^2}-\frac {m_s
v_1^2}{1+\sqrt{1-v_1^2}}+(2M_1-M_F)\sqrt{1-v_1^2}. \label{1.7}
\end{eqnarray}
Denoting
\begin{eqnarray}
w = \frac {v_1^2}{1+\sqrt{1-v_1^2}},\qquad U=(neV+\frac 1 2 m_s
v_0^2)c^{-2}, \label{1.7*}
\end{eqnarray}
and substituting it into (\ref{1.7}), we get
\begin{eqnarray}
w^2 M_1-\l[w+\ln({1-w})\r]M_F = ({1-w})U - w m_s.\label{1.8}
\end{eqnarray}
(\ref{1.8}) is a linear equation of $(M_1,M_F)$, which can be
easily solved by the method of least squares from a sequence of
measured data $(U_i, w_i),i=1,2,\cdots,N$. Define the
$N$-dimensional vectors  $(\vec a, \vec b, \vec f)$ by,
\begin{eqnarray}
\vec a&=& (w_1^2,w_2^2,\cdots, w_N^2),\\
\vec b&=& -\l[w_1+\ln({1-w_1}),w_2+\ln({1-w_2}),\cdots,
w_N+\ln({1-w_N})\r],\\
\vec f &=&[ ({1-w_1})U_1 -w_1 m_s, ({1-w_2})U_2 -w_2
m_s,\cdots,({1-w_N})U_N -w_N m_s].
\end{eqnarray}
Then the solution of the least square is given by
\begin{eqnarray}
M_1=\frac 1 D [\vec b~^2\vec a-(\vec a\cdot\vec b)\vec b]\cdot
\vec f,\quad M_F=\frac 1 D [\vec a^2\vec b-(\vec a\cdot\vec b)\vec
a]\cdot \vec f,\quad D = \vec b~^2 \vec a^2 -(\vec a\cdot\vec
b)^2,
\end{eqnarray}

For an electron, we have the typical order of magnitude for the
parameters in (\ref{1.8}),
\begin{eqnarray}
w\sim 1,\qquad U\sim m_s\sim 1{\rm MeV}.
\end{eqnarray}
So a meaningful test depends on the precision of the measurement
data $(w_i,U_i)$, which should be of relative errors less than
$10^{-4}$. How to promote the precision of the measurement is the
key for the success of a test.

Some possible solutions and its implications:
\begin{enumerate}
\item If $M_F=0$ and $M_1=0$, which means the Einstein's
mass-energy strictly holds, and the particles can not be described
by the classical fields, because a linear spinor is unstable for a
standing wave.

\item If $M_F=0$ and $M_1\ne 0$, this kind of particles has not
nonlinear self-potential, and the balance of the particles should be
explained by scalar and vectorial interactive fields.

\item If
$M_F \ne 0$, which means particle including nonlinear potential,
which leads to some unusual effects\cite{gu1,gu2,spn1}.
\end{enumerate}
so no matter what result the experiment provides, the implication
is always important and fundamental.

From the above analysis, we find the test of the energy-speed
relation can provide new insights into the structure of the
fundamental particles and space-time. An exact measurement of the
energy-speed relation is a shortcut to disclose the secrets of the
fundamental particles and  interactive potentials, as well as the
nature of dark matter and dark energy. So this experiment is of
fundamental significance, which may clarify a number of puzzles
 in physics.


\begin{thebibliography}{99}
\bibitem{E1} Einstein, A. Ann. der Phys. 18, 639-641 (1905).
\bibitem{test1}  S. Rainville, {\em et al., A Direct Test of $E = mc^2$}, Nature, 438,
22, 2005, p.1096.
\bibitem{test2} Ezzat G. Bakhoum, {\em Why $E = mc^2$ Emerges
in the Process of Neutron Capture}, arXiv:0705.3191
\bibitem{gu1} Y. Q. Gu, {\em New Approach to N-body Relativistic
Quantum Mechanics}, Int. J. Mod. Phys. A22:2007-2020(2007),
arXiv:hep-th/0610153
\bibitem{spn1} Y. Q. Gu, {\em A Cosmological Model with Dark Spinor
Source}, Int. J. Mod. Phys. A22:4667-4678(2007), arXiv:gr-qc/0610147
\bibitem{spn2} V. Adanhounme, A. Adomou, F. P. Codo, M. N. Hounkonnou, {\em Nonlinear spinor field equations in gravitational
theory: spherical symmetric soliton-like solutions}, arXiv:1211.3388
\bibitem{spn3} M. O. Ribas, {\em et al.}, Phys. Rev. {\bf D 72}, (2005)123502, arXiv:gr-qc/0511099.
\bibitem{spn4} H. J. de Vega, {\em Dark energy from cosmological neutrino condensation}, arXiv:astro-ph/0701212
\bibitem{spn5} S. I. Vacaru {\em Spinor and Twistor Geometry in Einstein
Gravity and Finsler Modifications}, Adv. Appl. Clifford Algebras 25
(2015) 453-483,
\bibitem{gu2} Y. Q. Gu, {\em Local Lorentz Transformation and Mass-Energy Relation of Spinor},
arXiv:hep-th/0701030v3

\end{thebibliography}
\end{document}